\documentclass{raa}            

\usepackage{graphicx,times}
\usepackage{natbib}
\usepackage{amssymb,amsmath}
\bibpunct{(}{)}{;}{a}{}{,}

\usepackage{multirow}   
\usepackage{makecell}
\usepackage{float}
\usepackage{verbatim}

\usepackage[a4paper=true,pagebackref=true,driverfallback=dvipdfm]{hyperref}
\hypersetup{colorlinks = true, linkcolor = green, anchorcolor = red, citecolor = blue, filecolor = red, pagecolor = red, urlcolor = red}

\begin{document}

   \title{Pulsar Candidate Classification Using A Computer Vision Method Combining with Convolution and Attention}

 \volnopage{ {\bf 20XX} Vol.\ {\bf X} No. {\bf XX}, 000--000}
   \setcounter{page}{1}

   \author{NanNan Cai
   \inst{1,2}, JinLin Han\inst{1,2,3}, WeiCong Jing\inst{1,2}, ZeKai Zhang\inst{4}, DeJiang Zhou
      \inst{1,2}, Xue Chen\inst{1,2}
   }

   \institute{National Astronomical Observatories, Chinese Academy of Sciences, 20A Datun Road, Chaoyang District, Beijing 100101, China; {\it nncai@nao.cas.cn}; {\it hjl@nao.cas.cn}\\
        \and School of Astronomy, University of Chinese Academy of Sciences, Beijing 100049, China
             \\
	\and CAS Key Laboratory of FAST, NAOC, Chinese Academy of Sciences, Beijing 100101, China \\
        \and Department of Physics, Boston College, Massachusetts 02467, USA \\
\vs \no
   {\small Received 20XX Month Day; accepted 20XX Month Day}
}

\abstract{Artificial intelligence methods are indispensable to identifying pulsars from large amounts of candidates.  
We develop a new pulsar identification system that utilizes the CoAtNet to score two-dimensional features of candidates, uses a multilayer perceptron to score one-dimensional features, and uses logistic regression to judge the scores above.
In the data preprocessing stage, we performed two feature fusions separately, one for one-dimensional features and the other for two-dimensional features, which are used as inputs for the multilayer perceptron and the CoAtNet respectively.
The newly developed system achieves 98.77\% recall, 1.07\% false positive rate and 98.85\% accuracy in our GPPS test set.
}

   \authorrunning{N.N. Cai et al. }            
   \titlerunning{Pulsar Candidate Classification}  
   \maketitle

%
\section{Introduction}           
\label{sect:intro}

Pulsars are known as rotating neutron stars whose radiation beam sweeps across the line of sight. The radio signals of these pulsars are pulsed in a wide radio band, but they are dispersed due to free electrons in the interstellar medium. Therefore, signals at a lower frequency are more delayed. The signals are received by a radio telescope and then converted to digital signals after a series of signal conversions and digital processes, finally are stored in a digital file.

\begin{figure*}
    \centering
    \includegraphics[width=1\textwidth]{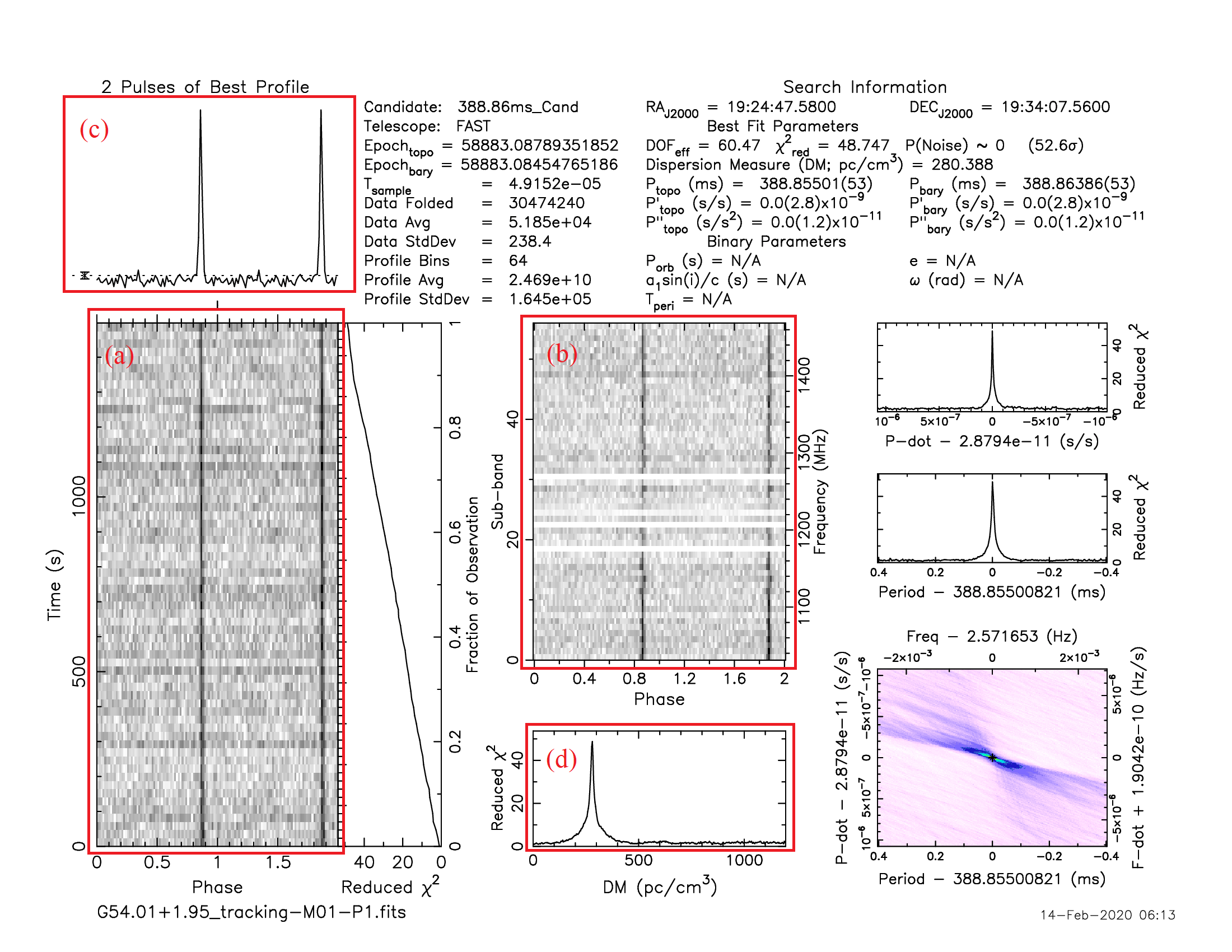}
    \caption{An example plot produced by the \texttt{prepfold} for a pulsar J1924+1932g discovered by the GPPS survey. Four key data sets are taken from the \texttt{pfd} file for the AI program: (a) the Time-vs-Phase plot for the integrated pulse strength over all frequency channels for many subintegration durations along time; (b) the Frequency-vs-Phase plot for the integrated pulse strength over all time against the many subbands in the observational band; (c) the pulsar pulse profile integrated over all time and frequency channels against the pulsar rotation phase; (d) the curve for the best DM shown by the reduced $\chi^2$ against DM.}
    \label{fig: pfdfile}
\end{figure*}

Astronomers recognize pulsar signals via searching the periodicity  ($P$) of pulses and the best dispersion measure ($DM$) of many trials for the delay compensation of pulses detected at different frequencies in the radio band. After data of many frequency channels in a radio band are de-dispersed (De-DM) and added together, one can find a possible period of a pulsar after the signals are analyzed via the Fast Fourier Transform (FFT) method. The analyses can be made in many packages. The most popular pulsar searching software is  PRESTO\footnote{\url{https://github.com/scottransom/presto}}. When  pulsar signals are recognized as a significant detection of the periodical signal from the long de-dispersed data set, the diagnostic plot can be produced via \texttt{prepfold}, the program for folding the original signals around the most probable period $P$ and the dispersion measure $DM$. 
Therefore, the folded data produced by \texttt{prepfold} are recorded in a \texttt{pfd} file with the features for diagnostics, which consists of one-dimensional features: the folded pulse profile based on the period and DM,  the change of signal-to-noise ratio (i.e., $\chi^2$) around the most probable period, the change of signal-to-noise ratio around the dispersion measure. There are also diagnostics expressed in the two-dimensions features: the detected signal in the time-versus-phase plot, or the frequency-versus-phase plot, or the signal-to-noise ratio over the P-versus-${\Dot{P}}$ plot, see Figure ~\ref{fig: pfdfile} for example.

Because of radio frequency interference (RFI), many such candidates are fake. In the past, astronomers had to pick out the true pulsars from all the candidates manually. With the development of new technique, the data volume on the pulsar search have become larger and larger, and pulsar candidates viewing and selection manually becomes more and more difficult. 

\begin{table*}
    \caption{Previous AI approaches for pulsar identification.  Note that the Recall, False positive rate (FPR), Accuracy, and F1 score are taken from the original references and should be interpreted with caution because they were obtained with different datasets.}
    \label{tab: Articles about pulsar identification.}
    \centering
    \begin{tabular}{llllll}
    \hline
    Method &   Recall & FPR  & Accuracy & F1 & Reference \\
    \hline
    12:12:2 ANN  & 92\% & - & - &- & \citet{Eatough2010}\\
    22:22:2 ANN  & 85\% & - & - & - & \citet{Bates2012}\\
    6:8:2 ANN  & 99\% & 0.11\% & - & - & \citet{Morello2014}\\
    SVM+ANN+CNN+LR & 92\% & 1\% & - & 96\% & \citet{Zhu2014} \\
    %
 GH-VFDT  & 92.8\%  & 0.5\%  & 98.8\%  & -  & \citet{Lyon2016} \\
          & 82.9\%  & 0.8\%  & 97.8\%  & -  &  \\
          & 78.9\%  & 0.1\%  & 99.8\%  & -  &  \\
    ensemble method with five Decision Tree & 98.7\% & 0.5\% & 99.2\% & - & \citet{Tan2018}\\
    SVM+ANN+ResNet+LR & 98\% & - & - & 92\% & \citet{Wang2019a} \\
    Random Forest+XGBoost & 96.7\%  & -  & 96.9\% & - & \citet{Wang2019b} \\
                          & 92\%    & -  & 91.8\% & - &  \\
    DCGAN+DeepF+SVM  & 100\% & 0\% & - & - & \citet{Guo2019}\\
    CCNN & 98.16\% & - & 94.76\% & - & \citet{Zeng2020} \\
    AdaBoost+Multi-input CNN & 95.6\% & - & - & 96.2\% & \citet{Lin2020a} \\
    GS+RFE+GBoost & 99\% & 0.102\% & - & - & \citet{Lin2020b} \\
    Semi-supervised GAN & 99.4\% & 1.6\% & 98.9\% & - & \citet{Balakrishnan2021} \\
    res-conv autoencoder & 99\% & - & 97.9\% & - & \citet{Yin2022a} \\
    Ada-GBoost-MICNN & 98.8\% & - & 98\% & - & \citet{Zhao2022} \\
    GAN+ResNeXt & 100\% & 0\% & 100\% & 100\% & \citet{Yin2022b} \\
    \hline
    \end{tabular}
\end{table*}

Previously, many methods have been developed to speed up the selection of candidates, such as Summary interfaces approaches and Semi-automated ranking approaches, which have been summary by \citet{Lyon2016}.
The identification of pulsars by using artificial intelligence (AI) is very efficient, see Table 1 for a list. 
\citet{Eatough2010} used a 12:12:2 artificial neural network (ANN) to evaluate twelve feature numbers of candidates. 
\citet{Bates2012} followed the approach and increased the feature number from 12 to 22.
A similar approach is used by \citet{Morello2014} for the High Time Resolution Universe (HTRU) survey. 
\citet{Zhu2014} combined four methods of machine learning, including Support Vector Machine (SVM), ANN, the Convolution Neural Network (CNN) and Logistic Regression (LR) to form the Pulsar Image-based Classification System (PICS) for recognizing pulse profile, DM-curve, time-versus-phase plot and frequency-versus-phase plot of pulsars. 
\citet{Wang2019a} changed the CNN module of PICS into a 15-layer residual CNN, which increases the depth of the network to improve accuracy and avoid the vanishing gradient caused by the increasing depth.
\citet{Tan2018} developed an ensemble classifier consisting of five different decision trees using mean value, standard deviation, skewness and excess kurtosis extracting from pulse profile, DM-curve, time-versus-phase, frequency-versus-phase.
It applied to radio pulsar search in the Low-Frequency Array (LOFAR) Tied-Array All-Sky Survey. 
A hybrid ensemble model combined Random Forest and XGBoost with EasyEnsemble was trained and tested on HTRU dataset \citep{Wang2019b}. 
\citet{Guo2019} employed the Deep Convolution Generative Adversarial Network (DCGAN) to generate more samples and learn features and adopt SVM to classify the candidates.
\citet{Lin2020a} designed a multi-input CNN which has a main input and an auxiliary input and utilized a Transforming Image and Adding Gaussian Noise (TIAGN) technique to augment data for solving the problem of training on a highly class-imbalanced data set.
\citet{Lin2020b} proposed the Grid Search and Recursive Feature Elimination for feature selection using the GBoost algorithm,  using 18 feature numbers for candidate classification.
\citet{Zeng2020} use the CNN to extract features from pulse profile, DM-curve, Frequency-versus-Phase plot and Time-versus-Phase plot, then concatenates these features horizontally or vertically for the input of the next layer of the network.
\citet{Balakrishnan2021} used pulse profile, DM-curve, Frequency-versus-Phase plot and Time-versus-Phase plot to train a semi-supervised generative adversarial network.
\citet{Yin2022a} proposed an interesting network of a residual convolutional autoencoder (RCAE) combined with LR. It only needs to train the Frequency-versus-Phase plots and Time-versus-Phase plots of non-pulsars.
\citet{Zhao2022} combined the AdaBoost with the multi-input-CNN (MICNN) to form a new framework, and he add a convolutional block attention module (CBAM) to the MICNN. It also uses four features: pulse profile, DM-curve, Frequency-versus-Phase plot and Time-versus-Phase plot. The Generative adversarial network and ResNeXt \citep{Xie2017} are applied to pulsar candidate selection and the precision, recall, and F1-score of 100\% are obtained on the HTRU Medlat data set.

In the history of artificial intelligence, convolutional architectures have always dominated the field of Computer Vision (CV) since the LeNet \citep{Yann1998}, AlexNet \citep{AlexKrizhevsky2012}, VGGNet \citep{Simonyan2015VeryDC} and other models have achieved great achievements. 
The Attention Mechanism has been extremely popular in the Natural Language Processing (NLP) since the Google Brain Team published their Transformer model \citep{Vaswani2017}.
Recently, the Attention Mechanism has also become a research hotspot in CV. What is the difference between CNN and Attention?
A convolutional layer has a convolution kernel with fixed values for each input, and the convolution kernel convolves with each local region of the input and outputs a Feature Map. 
In other words, the convolutional layer gives each local region of each input the same weight. In contrast, the Attention Mechanism is designed to give different parts of different inputs with different weights inspired by the human brain's tendency to focus on certain parts and ignore others when reasoning about information. 
There are three types of Attention Mechanisms in the CV: (1) Spatial attention mechanism, such as Non-local Attention \citep{Wang2018}, Spatial Transformer \citep{Jaderberg2015}; (2) Channel attention mechanism, such as Squeeze-and-Excitation Network (SENet) \citep{Hu2017}, Selective Kernel Network (SKNet)  \citep{Li2019}; (3) Mixed attention mechanism, such as CBAM \citep{Woo2018}, Residual Attention Network (ResAttNet) \citep{Wang2017}, CoAtNet \citep{Dai2021}. So far, the attention mechanisms have not yet been widely used to identify pulsar candidates.

The Five-hundred-meter Aperture Spherical radio Telescope (FAST) is the largest single-dish radio telescope in the world \citep{Nan2011}. We are carrying out a big pulsar searching project, namely the  FAST Galactic Plane Pulsar Snapshot survey \citet{Han2021}, and up to now, this project has discovered more than 500 pulsars\footnote{\url{http://zmtt.bao.ac.cn/GPPS/GPPSnewPSR.html}}.
In the data processing, millions of pulsar candidates are accumulated during the data processing. 
We take these as training and testing of new artificial intelligence methods of pulsars selection. 
We compared the effectiveness of three methods involving the Attention Mechanism in identifying two-dimensional features of pulsar candidates: 
Vision Transformer \citep{Alexander2021}, ResAttNet and CoAtNet. We find that the CoAtNet proposed by the Google Brain Team tends to perform better than others at discriminating the two-dimensional features.
Therefore the CoAtNet is chosen in this paper as the classifier for two-dimensional features of candidates. 

This paper combined the CoAtNet, MLP and LR to build our new pulsar identification system. 
The remaining part of the paper proceeds as follows: 
The overall structure of the pulsar identification system is introduced in Section 2. The setting and results of experiments about data pre-processing, selection of data set and the option of model parameters are presented in Section 3.
Finally, conclusions are given in Section 4.

\section{New approach for pulsar candidate identification}

\begin{figure*}
    \centering
    \includegraphics[width=1\textwidth]{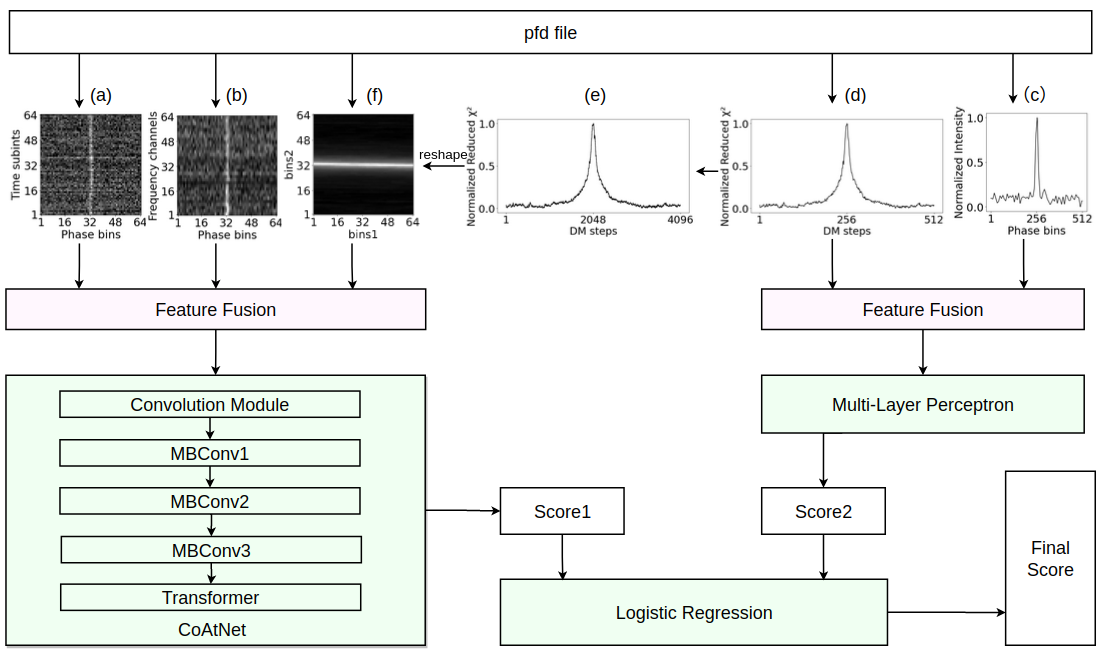} 
    \caption{The flow chart for the pulsar candidate selection and evaluation system. The inputs come from the \texttt{pfd} file, including (a) The $64\times64$ frequency-versus-phase plot; (b) The $64\times64$ time-versus-phase plot; (c) The normalized pulsar pulse profile over 512 bins; (d) The normalized reduced $\chi^2$ array for 512 DM steps. These datasets can have different sizes. Others are the converted datasets, such as (e) the normalized reduced $\chi^2$ array calculated for 4096 DM steps; (f) the plot is reshaped by the normalized reduced $\chi^2$ array which length 4096 bins but in two dimensions. The features are fused and fed to the CoAtNet and MultiLayer Perceptron, and the outputs are further evaluated in the Logistic Regression for the final score.}
    \label{fig: system structure}
\end{figure*}

The flow chart of our newly designed system for pulsar candidate identification using the machine learning method is presented in Figure~\ref{fig: system structure}.

We extract two-dimensional features and one-dimensional features from a \texttt{pfd} file for pulsar candidates for Pulsar candidate identification. Feature fusion is used to combine multiple feature information into a feature. Two feature fusions have been made in our method. First, the feature fusion has been carried out for 3 inputs of 2-D image features (a, b, f), and the combined feature is fed into CoAtNet; Second, feature fusion has been carried out for 2 inputs of 1-D features (d, c), and the combined feature is fed into MultiLayer Perceptron (MLP). 

Then, we take the outputs of the CoAtNet and MLP as the inputs of the model implemented using the LR method. Our method is named as ``CoAtNet-MLP-LR''. 
The specific details of data processing by using ``CoAtNet-MLP-LR'' are introduced in Section 3. In this Section, we want to describe the network architecture and theory of CoAtNet, MLP and Logistic Regression.

\subsection{CoAtNet}

In the field of machine learning, the capacity and generalization ability of a model are critical factors for its success. The model capacity of a machine learning model refers to its ability to learn complex patterns from data, while the generalization ability refers to how well the model performs on unseen data. A critical issue for a model to have high generalization capacity is to avoid overfitting, a phenomenon in which a model memorizes the training data, leading to poor performance on new, unseen data.
Google Brain’s study \citep{Dai2021} concluded that the convolutional module has better generalization, while the attention module has higher model capacity.
To address this trade-off between model capacity and generalization, they proposed a novel model called CoAtNet, which marries the depthwise convolution and attention mechanism. By combining the generalization capacity of depthwise convolution with the model capacity of Transformer, CoAtNet is expected to achieve better performance on both capacity and generalization.

\begin{figure*}
    \centering
    \includegraphics[width=0.8\textwidth]{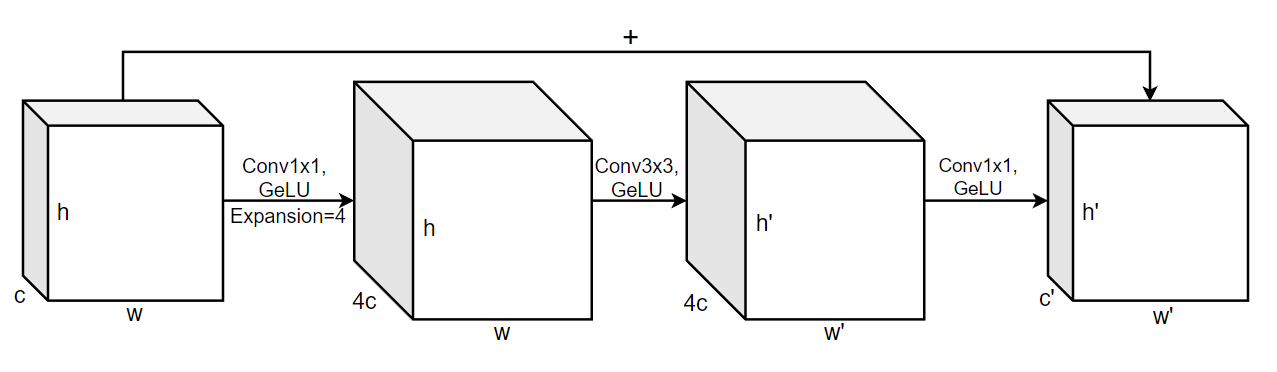}
    \caption{Detailed operation in the MBConv block.}
    \label{fig: MBConv}
\end{figure*}

The depthwise convolution block proposed by MobileNetV2 \citep{Sandler2018} -- MBConv is shown in Figure \ref{fig: MBConv}, which includes three steps: (1) Expanding the channels of input using convolution with kernel size $1\times1$; (2) applying a  Convolution with kernel size $3\times3$ and pad 2; (3) compressing output channels using a Convolution with kernel size $1\times1$.
Each step is followed by an activation function, where GeLU\citep{Hendrycks2016} is used in our work. 
Depthwise convolution has a lower computational cost and smaller parameter size, and its inverted residual bottleneck design helps improve recognition accuracy.

In the Transformer block, CoAtNet replaces the self-attention mechanism with the relative-attention mechanism, which is a natural mixture of depthwise convolution and attention with minimum additional cost.
The self-attention mechanism is the core of the famous Transformer model. It trained matrix $W_Q$,$W_K$,$W_V$ and multiplied the input X to get matrix Q, K, V.
Then it realizes "dynamic weight" for different input X by the following formula:
\begin{equation}
    Y = softmax( \frac{QK^T}{\sqrt{d_K}} )V.
\end{equation}

CoAtNet is a deep neural network architecture comprising five vertically arranged stages, denoted by $S0$, $S1$, $S2$, $S3$, and $S4$. While $S0$ uses a standard convolution block, $S1$ to $S4$ can adopt either an MBConv block or a Transformer block, with the former always preceding the latter. 
As such, CoAtNet offers five possible variants: (1) C-C-C-C; (2) C-C-C-T; (3) C-C-T-T; (4) C-T-T-T; (5) T-T-T-T (i.e.${\rm ViT_{REL}}$).
According to Google Brain’s experiment, the variants can be ranked as follows by model capacity: C-C-T-T $\approx$ C-T-T-T $>$ ${\rm ViT_{REL}}$ $>$ C-C-C-T $>$ C-C-C-C; for the generalization, the ranking is: C-C-C-C $\approx$ C-C-C-T $\ge$ C-C-T-T $>$ C-T-T-T $\gg$ ${\rm ViT_{REL}}$.
 To achieve a balance between model capacity and generalization capacity, we chose to experiment with the C-C-C-T, C-C-T-T, and C-T-T-T variants on our dataset. Our experimental results indicate that the C-C-C-T variant is the most effective, as further detailed in Section 3.

\subsection{Multi-layer Perceptron}

A multi-layer perceptron is a type of fully connected feedforward artificial neural network that consists of an input layer, an output layer, and one or more hidden layers. The input layer receives the input features, and each feature is connected to each artificial neuron node in the first hidden layer, resulting in a fully connected network.
If we assume that the input feature is $X=[x_1, x_2, ..., x_n]$ and the first hidden layer has $m$ artificial neuron node $(a_1, a_2, ..., a_m)$.
The output of the input feature X after passing through the $j$th neural node is:
\begin{equation}
    h_j=g(\sum_{i=1}^{n}w_i{}_j \cdot x_i + b_j)
\end{equation}
Here, function $g$ is the activation function, which introduces non-linearity and enhances the learning ability of the neural network.
The parameter $w_i{}_j$, $b_j$are learned by the backpropagation algorithm during the training process.
In our work, we use the Log Loss Function, i.e., Cross-Entropy Loss Function as the loss function.
Then, the gradient of the loss with respect to each parameter can be calculated by the chain rule.
For instance, in a simple MLP with one hidden layer, the outputs of the hidden layer $H$ and the output layer $O$ can be computed as follows:
\begin{equation}
    H = g^{(1)}(W^{(1)} \cdot X + B^{(1)})
\end{equation}
\begin{equation}
    O = g^{(2)}(W^{(2)} \cdot H + B^{(2)})
\end{equation}
If the loss between the predicted output $O$ and true labels $Y$ is denoted by $L(Y,O)$, the gradient of $W$ can be computed as follows:
\begin{equation}
    \Delta{W^{(2)}} = \frac{\partial L}{\partial O} \cdot \dot g^{(2)} \cdot H
\end{equation}
\begin{equation}
    \Delta{W^{(1)}} = \frac{\partial L}{\partial O} \cdot \dot g^{(2)} \cdot W^{(2)} \cdot \dot g^{(1)} \cdot X
\end{equation}

The neural network parameters are iteratively updated based on their gradients, allowing for fast and nonlinear learning.
While the prediction accuracy of neural networks is generally high, it may be slightly lower than that of more advanced methods such as Support Vector Machines, which require more computationally expensive matrix calculations. In our pulsar identification task, we found that the Frequency-versus-Phase plot was the most important one-dimensional feature. To identify this feature, we employed the simple and fast MLP algorithm. However, for two-dimensional features, we used a relatively more complex algorithm. Specifically, the MLP block in our pulsar candidate identification system consisted of four hidden layers, with 2048, 4096, 1365, and 455 neuron nodes, respectively.
The MLP block in the pulsar candidate identification system has four hidden layers, the number of neuron nodes in each layer is 2048, 4096, 1365 and 455, respectively.

\subsection{Logistic Regression}

Logistic Regression is a classification algorithm that predicts the probability of an event taking place.
For the case of one or more independent variables and a single binary dependent variable, the model predicts the probability of the positive class as:
\begin{equation}
    p(y_i=1|X_i) = \frac{1}{1+e^{-(W \dot X_i + W_0)}}
\end{equation}
The parameter is estimated using maximum likelihood estimation, and it is not possible to find a closed-form solution that maximizes the likelihood function, so that must be searched using an iterative process.
The goal that maximizing the log-likelihood $\ln{L(W)} = \sum [y_i \ln{p_i} + (1-y_i) \ln{(1-p_i)}]$ is equivalent to that minimizing the cost function with regularization term $r(W)$:
\begin{equation}
    C \sum_{i=0}^n[-y_i\ln{p_i}-(1-y_i)\ln{(1-p_i)}] + r(W)
\end{equation}

\section{Experiments and Results}

\subsection{Data Set}

The FAST Galactic Plane Pulsar Snapshot survey project has accumulated tens of thousands of candidates including pulsars, their harmonics and fake candidates. We select some of them to form a data set, which is divided into a training data set and a testing data set. 
The training data set determines the upper limit of how well the model can learn, and the test data set is the standard to judge the performance of the model. The ratio of positive and negative samples (i.e. true pulsars and RFI) in all our training sets is approximately 1:1.

Experiments show that the data sets for the model are very important, which must be large enough and include different types of real pulsars including drifting, nulling, scattering, binary, intrinsic wide pulse and so on. The detected pulsar harmonics with multiple peaks are excluded from the data set. 
Include some pulsars with multiple peaks in the data set to ensure this type of pulsar can be identified.

A model with higher model capacity can capture more details of features in training data, but it is easier to overfit in the case of an insufficient amount of training data set. As a result, the generalization ability of the model is worse (i.e. the model performs well on the training data set but not on the testing data set).
We experimented with training data of different sizes (2000, 6000, 8000, 12000, 16000), in order to find out how much data can play the model capacity of our model. 
The experimental results are shown in Table \ref{tab: Comparison of training results with different data amounts}.

The inputs of our pulsar candidate identification system are the pulse profile, reduced $\chi^2$ values of DMs, frequency-versus-phase plot and time-versus-phase plot.
Human experts identify pulsars based on the pulsar diagnostic plots and the above features are mainly taken into account.
These features are extracted from pfd file, and the pulse profiles are normalized and resized to 512 bins, the reduced $\chi^2$ values of DMs are calculated to 512 bins and normalized, the frequency-versus-phase plots and the time-versus-phase plots are reshaped to $64\times64$ 2-D arrays.

Past research using AI methods to identify pulsars usually used a separate model to judge each feature individually.
We experimented with feature fusions in the stage of data preprocessing and obtained better results than the results of judgment with independent features.
The purpose of feature fusion is to combine multiple feature information into a feature that is easier to distinguish than the original feature.
The feature fusions work as follows: 
When pulsar profile data are extracted, normalized and resized to 512 bins, it will combine with the normalized reduced $\chi^2$ array length 512 to form a 1-D merged array length 1024. 
We also calculate the reduced $\chi^2$ values to 4096 bins and reshape it to a $64\times64$ 2-D array.
Then the frequency-versus-phase array, the time-versus-phase array and the 2-D array of reduced $\chi^2$ values are combined to a "multi-channel image", each two-dimensional feature is regarded as a "single channel image", which, like the RGB image consists of three channels of red, green and blue.

Without feature fusion, multiple features are predicted separately, and then the predicted results are synthesized to form the final predicted score.
But in our system, the ANN model can capture the relationship between profile and reduced $\chi^2$ values, and the CoAtNet model can capture the relationship between the frequency-versus-phase plot, the time-versus-phase plot and the 2-D array of reduced $\chi^2$ values.
In fact, it's easy to imagine that judgment with feature fusion is closer to human judgment thinking.
Human experts often combine details of the frequency-versus-phase plot, the time-versus-phase plot and the DM-curve on the diagnostic plots to determine whether a candidate is a pulsar.
In order to verify the effectiveness of feature fusion, a comparative experiments are conducted, and the experimental results are shown in Table \ref{tab: feature fusion}.

\subsection{Performance Evaluation}
Recall, false positive rate(FPR) and accuracy are usually used to evaluate the performance of models for machine learning classification tasks.
Recall is the ratio of true positives (TP) to the sum of true positives and false negatives (FN):
\begin{equation}
    Recall = \frac{TP}{TP+FN}
\end{equation}
While FPR is the ratio of false positives (FP) to the sum of false positives and true negatives (TN):
\begin{equation}
    FPR = \frac{FP}{FP+TN} 
\end{equation}
Accuracy is the ratio of the sum of true positives and true negatives to the sum of true positives, true negatives, false positives, and false negatives (FN):
\begin{equation}
    Accuracy = \frac{TP+TN}{TP+TN+FP+FN}
\end{equation}

In our experiments, we define recall as the ratio of correctly identified pulsars to the total number of true pulsars in the dataset.
The false positive rate refers to the ratio of the number of false candidates identified as pulsars to the total number of false candidates in data set.
Accuracy is the proportion of all candidates correctly identified in the dataset to the total number of candidates.
The training goal of our AI model is to maximize recall while minimizing the FPR.

It is worth noting that our recognition system produces a score between 0 and 1, with a candidate being classified as a pulsar if its score exceeds a threshold value. 
In all experiments of this paper, we set the threshold value at 0.5. 
However, to minimize the risk of missing any potential pulsars, a more conservative threshold of 0.1 is adopted in the practical application of the GPPS project.
Any candidate with a score greater than 0.1 is classified as a pulsar and subsequently undergoes manual inspection.

\subsubsection{Finding The Most suitable variant}

From section 2, we know that the MBConv block or Transformer block can be used for the four stages of CoAtNet, so there are five CoAtNet variants: C-C-C-C, C-C-C-T, C-C-T-T, C-T-T-T, T-T-T-T.
Considering the balance of the model ability and generalization ability, we choose the C-C-C-T variant, C-C-T-T variant and C-T-T-T variant to experiment on the fused two-dimensional feature in our data set.
These models were trained on a training set that consisted of 6333 pulsar samples and 6148 RFI samples, and their parameters were saved for testing. We evaluated these models on a test dataset containing diverse types of pulsar and RFI samples, and their performance metrics are shown in Table~\ref{tab: The comparison of different variants}, each performance metric of the C-C-C-T CoAtNet performed better than other variants.
It's indicated that the CoAtNet with C-C-C-T architecture is the most suitable variant for our data, so we use it to build our pulsar identification system. 
\begin{table}
    \caption{The performance on the Recall, False Positive Rate and Accuracy Rate for three CoAtNet variants, trained using the same data set. The CoAtNet variant C-C-C-T is the best.} 
    \label{tab: The comparison of different variants}
    \centering
    \begin{tabular}{cccc}
     \hline
     & Recall & False Positive Rate & Accuracy \\
     \hline
     C-C-C-T & \textbf{98.75\%} & \textbf{1.31\%} & \textbf{98.72\%} \\
     C-C-T-T & 98.15\% & 2.99\% & 97.61\% \\
     C-T-T-T & 94.84\% & 4.29\% & 95.26\% \\
     \hline
    \end{tabular}
\end{table}

\subsubsection{The model capacity of our model}

We trained five models using five training datasets with sample sizes of 2000, 6000, 8000, 12000, and 16000, and then use the same test dataset to evaluate their performance.
To assess the performance of these models, we utilized a test dataset consisting of 6320 true pulsar samples and 5717 RFI samples, and measured their recall, false positive rate, and accuracy.
As shown in Table~\ref{tab: Comparison of training results with different data amounts}, increasing the sample size in the training dataset improved the model's performance, with the best accuracy reaching about 98.85\% on the GPPS test dataset.
However, we observed that the model's accuracy tends to plateau when the training dataset size exceeds 12000, indicating that the proposed model requires a substantial amount of input data, with a minimum of 12000 samples necessary to achieve optimal performance.

\begin{figure}
    \centering
    \includegraphics[scale=0.35]{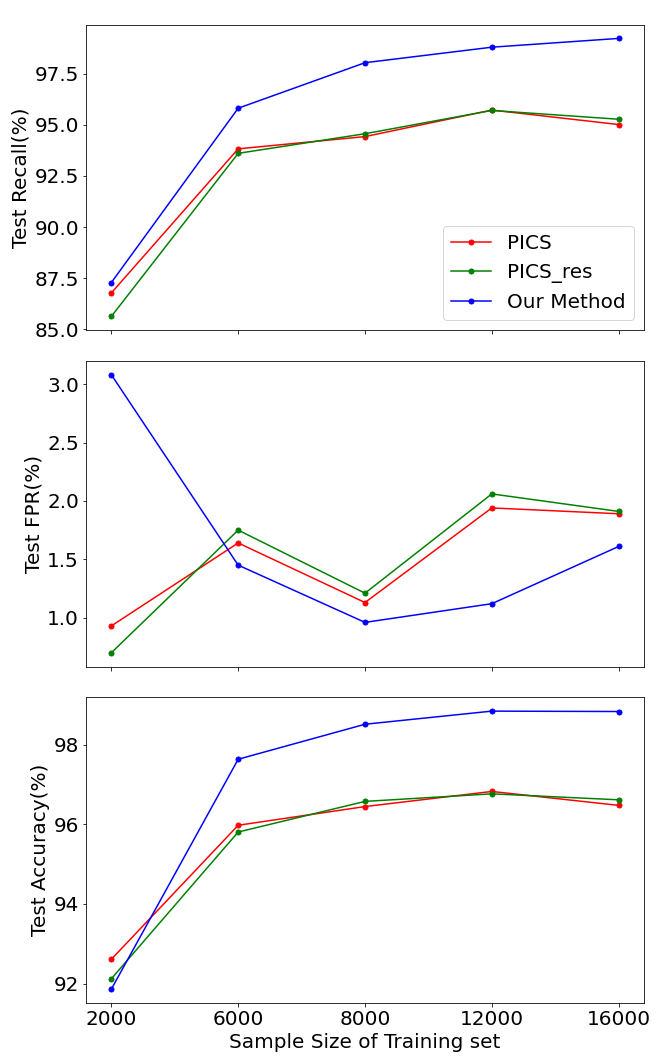}
    \caption{The recall, FPR and accuracy in test data set of PICS, PICS\_res and our method using five training sets of different size}
    \label{fig: Comparison of results with different data amounts}
\end{figure}

To compare the proposed method with existing methods, we trained two other methods, namely PICS and PICS\_res, using the same training datasets with different sample sizes and evaluated their performance on the test dataset.
Figure~\ref{fig: Comparison of results with different data amounts} shows that the proposed method outperformed the other two methods in terms of identification accuracy, especially when trained on a sufficient amount of data.
Our experimental results demonstrate that as the size of the training dataset increases, the proposed method outperforms the other two methods, indicating its stronger model capacity. Specifically, when the amount of training data is increased by the same amount, our method exhibits a faster increase in accuracy compared to the other two methods. These findings suggest that the proposed method can effectively benefit from a larger training dataset and has a higher potential for improving performance with additional training data. In summary, our results indicate that sufficient training data is crucial for achieving high identification accuracy, and the proposed method has a stronger capacity to utilize larger training datasets than the other two methods.
\begin{table}
 \centering
 \caption{Performance metrics for different sizes of training samples}
 \label{tab: Comparison of training results with different data amounts}
 \setlength{\tabcolsep}{4.0pt}
 \footnotesize
 \begin{tabular}{l|c|c|c|c|c}
  \hline
  Size of training sample & ~2000 & ~6000 & ~8000 & ~12000 & ~16000\\
  \hline
  Recall in training(\%) & 99.36 & 99.97 & 99.92 & 100 & 100\\
  Recall in testing(\%) & 87.28 & 96.82 & 98.05 & 98.81 & 99.24\\
  \hline
  FPR in training(\%) & 2.96 & 0.0 & 0.12 & 0.0 & 0.0\\
  FPR in testing(\%) & 3.08 & 1.45 & 0.96 & 1.12 & 1.61\\
  \hline 
  Accuracy in training(\%) & 98.19 & 99.98 & 99.9 & 100 &  100\\
  Accuracy in testing(\%) & 91.86 & 97.64 & 98.52 & 98.85 & 98.84\\
  \hline
 \end{tabular}
\end{table}

\subsubsection{The Usefulness of feature fusion}
We conducted a pair of control experiments to demonstrate the effectiveness of feature fusion.
The experiment \uppercase\expandafter{\romannumeral1} involved training five models using the pulse profile 1-D array length 512 bins, the reduced $\chi^2$ values 1-D array length 512 bins, the $64\times64$ frequency-versus-phase 2-D array, the $64\times64$ time-versus-phase 2-D array and the $64\times64$ reduced $\chi^2$ values 2-D array. 
The predicted results of these models were then synthesized using LR.
In experiment \uppercase\expandafter{\romannumeral2}, we trained ANN models using the combined 1-D feature, trained the CoAtNet model using the combined 2-D feature, and synthesized the predicted results of the two models using LR.
The identification results with and without feature fusion are presented in Table~\ref{tab: feature fusion}, demonstrating an increase in the final test accuracy from $97.68\%$ to $98.85\%$ with feature fusion.

\begin{table}
    \centering
    \caption{The accuracy for pulsar candidate identification by using fused features or individual features via methods in ``CoAtNet-MLP-LR''. Feature fusion can give a higher accuracy.} 
    \label{tab: feature fusion}
    \begin{tabular}{lll}
    \hline
     Input features & Method & Accuracy  \\
     \hline
     (c): 1-D profile   & MLP & 88.62\% \\
     (d): 1-D DM curve  & MLP & 92.96\% \\
    {\bf (c)+(d): fused 1-D feature} & MLP & {\bf 95.33\%} \\
     \hline
     (b): 2-D frequency-phase plot & CoAtNet & 98.31 \% \\
     (a): 2-D time-phase plot      & CoAtNet & 91.90\%  \\
     (f): 2-D DM plot              & CoAtNet & 92.81\%  \\
    {\bf  (a)+(b)+(f): fused 2-D feature }& CoAtNet & {\bf 98.47\%}  \\
     \hline
     (a)$\,\to\,$CoAtNet;  (b)$\,\to\,$CoAtNet; (f)$\,\to\,$CoAtNet; (c)$\,\to\,$MLP;  (d)$\,\to\,$MLP & LR & 97.68\% \\
     {\bf (c)+(d)$\,\to\,$MLP ; (a)+(b)+(f)$\,\to\,$CoAtNet}  & LR & {\bf 98.85\% }  \\
     \hline
    \end{tabular}
\end{table}

\subsection{Comparison with Other Methods}
We have searched all the papers in Table ~\ref{tab: Articles about pulsar identification.}.
Among them, the methods proposed by  \citet{Eatough2010},  \citet{Bates2012} and  \citet{Morello2014} were found to be easily reproducible but outdated.
And in the remaining papers, these methods PICS \citep{Zhu2014}, PICS\_res \citep{Wang2019a}, CCNN \citep{Zeng2020} and SGAN \citet{Balakrishnan2021} are provided with codes. 
The code of PICS AI and its upgrade PICS\_res can be downloaded from \url{https://github.com/zhuww/ubc\_AI}, and they are easy to train using our own training set.
The code of CCNN and SGAN can be downloaded from \url{https://github.com/xrli/CCNN/} and \url{https://github.com/vishnubk/sgan} respectively, but we met some difficulties when re-training our training data set.

To ensure a valid and meaningful comparison between two or more methods, it is imperative to train them using the same training set and evaluate their performance on the same test set. In our comparative analysis, the re-trained PICS and re-trained PICS\_res have undergone this type of rigorous evaluation.
And we use the GPPS test dataset to test the re-trained PICS model and the re-trained PICS\_res model, and also test the re-trained PICS model and the re-trained PICS\_res model. 
The assessment result of our method and other methods are shown in Table~\ref{tab: The comparison with other methods}.
So the comparison between these retrained models and our approach is fair and valid, and our method achieves 98.77\% recall, 1.07\% false positive rate and 98.85\% accuracy; it gets better on each performance evaluation metric.
The models without re-trained here are for reference only.

\begin{table}
    \centering
    \caption{Performance metrics of our method compared with those of other methods, which are evaluated by using the same test data set. '*' indicates a re-trained model using our data.}
    \label{tab: The comparison with other methods}
    \begin{tabular}{ccccl}
        \hline
         & Recall & False Positive Rate & Accuracy & Reference\\
        \hline 
      {\bf   CoAtNet-MLP-LR} & 98.77\% & 1.07\% & 98.85\%  & this work\\[1mm]
        PICS & 86.77\% & 0.93\% & 92.61\%  & \cite{Zhu2014}\\
        PICS* & 95.73\% & 1.94\% & 96.83\% &  \cite{Zhu2014}\\
        PICS\_res & 72.17\% & 4.78\% & 83.12\%  & \cite{Wang2019a}\\
        PICS\_res* & 95.71\% & 2.06\% & 96.77\%  & \cite{Wang2019a}\\
        CCNN & 96.28\% & 7.17\% & 94.65\% &  \cite{Zeng2020}\\
        SGAN & 67.43\% & 4.90\% & 80.58\% & \cite{Balakrishnan2021}\\
        \hline
    \end{tabular}
\end{table}

\section{Conclusions}
We applied CoAtNet, a new deep learning algorithm for image classification, and developed  ``CoAtNet-MLP-LR'' for pulsar candidate identification, combined with the feature fusion method. Then, larger training and testing sets are selected and a series of experiments are conducted, and it is found that the C-C-C-T CoAtNet is more suitable for pulsar identification. Our method ``CoAtNet-MLP-LR'' has a stronger model capacity and requires a larger amount of training data to give full play to its model capacity, the training data set should contain about 10,000 samples. 
Based on a large amount of training data set, we get a well-trained result with about 98\% accuracy. 
Our feature fusion plan is useful in pulsar identification, which increases the accuracy from 97.68\% to 98.85\%.
Through the results of the comparative experiments in Table~\ref{tab: The comparison with other methods}, our model performs best on the test set of GPPS project, and the identification accuracy can reach 98.85\%, proving that it is superior to other methods.

\section{Code and Data Available}
The code of our pulsar identification system ``CoAtNet-MLP-LR'' is available in a GitHub repository \footnote{\url{https://github.com/cnn-OvO/pulsar-identification.git}}.
If you would like to get the training data or have any questions, please contact us at \url{nncai@nao.cas.cn}.

\normalem
\begin{acknowledgements}
This work is supported by the National Natural Science Foundation of
China (NSFC, Nos. 11988101 and 11833009) and the Key Research 
Program of the Chinese Academy of Sciences (Grant No. QYZDJ-SSW-SLH021).
\end{acknowledgements}
  
\bibliographystyle{raa}
\bibliography{bibtex}

\end{document}